\documentclass[authoryear]{sigplanconf}

\pdfoutput=1
\usepackage{url}
\usepackage{amsmath}
\usepackage{natbib}
\usepackage{graphicx}
\usepackage{booktabs}
\usepackage{subfigure}
\usepackage[colorlinks,
            linkcolor=blue,
            anchorcolor=blue,
            citecolor=blue]{hyperref}

\begin{document}

\setlength{\pdfpageheight}{\paperheight}
\setlength{\pdfpagewidth}{\paperwidth}





\titlebanner{banner above paper title}        
\preprintfooter{short description of paper}   

\title{Rethinking complexity for software code structures: \\ A pioneering study on Linux kernel code repository}

\authorinfo{Wenhe Zhang}
           {Peking University}
           {wenhe@pku.edu.cn}

\authorinfo{Jin He}
           {VMware(China) Inc.}
           {hejin@vmware.com}

\authorinfo{Kevin Song}
           {VMware(China) Inc.}
           {skevin@vmware.com}

\maketitle

\begin{abstract}
The recent progress of artificial intelligence(AI) has shown great potentials for alleviating human burden in various complex tasks. From the view of software engineering, AI techniques can be seen in many fundamental aspects of development, such as source code comprehension, in which state-of-the-art models are implemented to extract and express the meaning of code snippets automatically. However, such technologies are still struggling to tackle and comprehend the complex structures within industrial code, thus far from real-world applications. In the present work, we built an innovative and systematical framework, emphasizing the problem of complexity in code comprehension and further software engineering. Upon automatic data collection from the latest Linux kernel source code, we modeled code structures as complex networks through token extraction and relation parsing. Comprehensive analysis of complexity further revealed the density and scale of network-based code representations. Our work constructed the first large-scale dataset from industrial-strength software code for downstream software engineering tasks including code comprehension, and incorporated complex network theory into code-level investigations of software development for the first time. In the longer term, the proposed methodology could play significant roles in the entire software engineering process, powering software design, coding, debugging, testing, and sustaining by redefining and embracing complexity.

\end{abstract}



 \\
  \indent\indent\indent \quad  complex network theory, knowledge representation.

\section{Introduction} 
\subsection{Research Problem Scoping}
The past decade witnessed artificial intelligence's growing power in almost all walks of life. From visual objects to natural languages, the target data for AI models to learn has become more abstract, complex, and structured \cite{2016Building}. With this trend, the idea of training models to comprehend code, a highly abstract form of language with rich compositional structures, naturally emerged in AI research. Learning and representing the semantics, rules, and structures in programs are regarded as a significant indicator of building more robust AI \cite{allamanis2018survey}.

\vspace{0.5em}
\noindent The recent years have also brought about more scientific findings on the process of software engineering, as it is an indispensable factor for high-tech companies' profits. In an array of research work, the roles of code in software engineering were highlighted. For engineers, reading code and searching for related information occupy about 50\% of their working time \cite{maalej2014comprehension}. On the long timeline of software development, a considerable amount of time and efforts will be spent on code repair, performance optimization, and requirements tracking\cite{Amy2016coop}. The cost of code-related tasks stimulated the increasing need for automatic code comprehension in the industry.

\vspace{0.5em}
\noindent As academic proceedings merge with real-world needs, a practical issue is the introduction of AI-based code comprehension into industrial settings. The software source code is usually large-scale and complicated, posing unprecedented challenges for current AI models: How to represent and process the source code of large industrial projects? How to evaluate the difficulty and performance of real-world code comprehension? How to deal with the complexity of software source code? Our work starts from the long-standing Linux kernel source code, towards relevant solutions by dataset collection, representation construction, and network analysis.

\subsection{The Loss of Current AI-based Code Comprehension}
\noindent On the basis of above problems, the related progress of AI technology has been motivating us to dig deeper. Aware of the analogy between natural languages and programming languages, many researchers recently proposed to tackle the code comprehension problems by natural language processing (NLP) approaches \cite{allamanis2018survey,sun2014empirical}. Through encoding methods or deep learning models for embedding, code snippets were transformed into distributed vector representations and fed into models or frameworks that are previously designed for NLP tasks \cite{chen2019literature}. The performances of such pipelines on code processing tasks were impressive. With vector representations, the encoder-decoder framework could help provide feedback for students' coding assignments \cite{piech2015learning}; transformer-based models achieved state-of-the-art performance in the summarization tasks of Python and Java code \cite{ahmad2020transformer}; the dual learning framework was powerful in both describing code snippets by natural language and generating code snippets from natural language descriptions \cite{wei2019code}. 

\vspace{0.5em}
\noindent However, NLP technology's success in processing code snippets does not mean these models are fully capable of code comprehension. While treating code as natural language brings convenience for data processing, it leads to the overdependence on relatively shallow semantic information, such as the naming of variables and functions in code. It also results in the neglect of critical structural information, including recursive calls, control structures, and data flow \cite{nguyen2013statistical, allamanis2015bimodal, ben2018neural}. Admittedly, this methodology can achieve satisfactory results when dealing with programming languages centered on scripts and glued by a bunch of commands, such as Python and SQL \cite{yao2018staqc}. But it still struggles to comprehend more abstract languages, including C and C++, in which complex program structures and call relations are of more importance. As these programming languages are the foundations of applications related to operating systems, databases, compilers, and so on, the targeted comprehension is an increasingly pressing issue for many high-tech companies.

\vspace{0.5em}
\noindent Digging into this phenomenon, we discovered that the capacity and limitations of current AI-based code comprehension are mainly shaped by the code datasets used in model training:

\begin{itemize}
    \item From the aspect of data resources, many of the datasets were collected from websites containing open-source code, such as Github and StackOverflow \cite{wong2015clocom, yao2018staqc}. Through web information extraction, massive code data and related annotations were included in a dataset, providing adequate labeled materials for training AI models. However, the code samples in such datasets are usually short, simple, and mutually independent snippets. It is hard to generalize the model trained on these datasets to our industrial software code, which is characterized by long-range dependencies and complex structures, as well as interleaved relations between different code blocks.
    \item From the aspect of data preprocessing, the characters related to the syntax of programming languages were excluded from the code snippets in datasets \cite{sun2014empirical}. For example, when preprocessing Python code, indentations in the code were utterly removed \cite{ahmad2020transformer}. While such a data preprocessing method simplifies the subsequent modeling procedure, it discards the structural information within the code, "flattening" the compositional structures into token sequences.
\end{itemize}

\vspace{0.5em}
\noindent The constraints from data hindered the development of higher-level automatic systems for software engineering tasks based on complex code comprehension, including requirement tracking \cite{liou2011toward}, code repair \cite{devlin2017semantic}, and error location \cite{lam2017bug}. If datasets closer to real-world software source code are created, we will directly approach possible solutions to the complexity problem in code comprehension and software engineering.

\subsection{Linux Kernel Source Code as Data Resource}
 When it comes to dataset construction on code with industrial strength, Linux kernel source code stands out as a perfect choice. Its repository scale has shown a considerable potential to serve as the data resource for a set of software engineering tasks.  According to our statistics, there are currently 29079 C files in the open-source repository of Linux kernel (version 5.8.11) and 15 to 16 functions in each file. The characteristics of the repository are also promising for data collection:
\begin{itemize}
    \item Firstly, it is a \emph{self-contained} repository without dependence on any external knowledge resources, as an advantageous feature for efficient data management and knowledge mining.
    \item It also possesses a large amount of \emph{labeled} code data. The abundant comments in the source code have provided various labels, including function summaries, input/output descriptions, and indications of important steps within functions.
    \item Moreover, the code blocks in Linux kernel repository are \emph{interrelated}, linking with each other by a series of call relations. This feature serves as the prerequisite for exploring complexity problems on source code.
\end{itemize}

\vspace{0.5em}
\noindent With the above considerations, we performed a systematical dataset collection work on the latest open-source repository of Linux kernel. Through our work, we hope to provide a data foundation for tackling complex code structures in large-scale software code, and powering software engineering by reliable code comprehension in real-world settings.

\subsection{Thinking Complexity in Software Engineering}

The phenomena of complexity are common and long-existing in the process of software engineering, including design, cooperation, management, and so on \cite{Amy2016coop}. As an intuitive solution, complex network theories were extensively applied to represent these problems \cite{wen2007software}. For example, the related work has modeled the collaboration of software engineering \cite{myers2003software}, the execution of software\cite{cai2009software}, and the relations between software packages \cite{zheng2008analyzing} as complex networks, providing tremendous insights and convenience for the industry's software development efforts.

\begin{figure}
\centering
\includegraphics[width=0.90\linewidth]{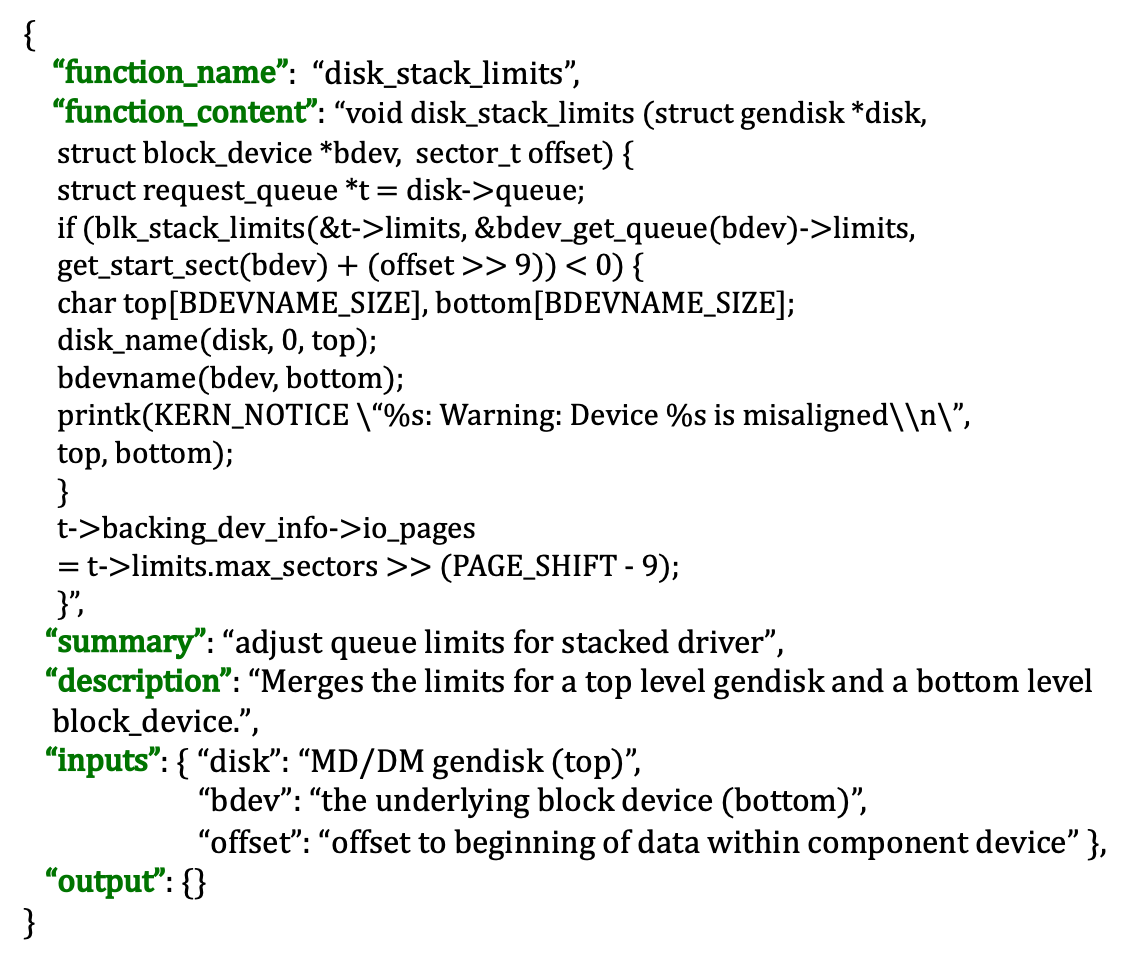}
\nocaptionrule
\caption{\textbf{An example sample in our present dataset}}
\label{fig1}
\end{figure}

\vspace{0.5em}
\noindent However, these network representations were mainly targeted at the high-level processes of software engineering, without naturally extending them to the modeling of source code, which is the most fundamental component for software and the origin of almost all the complexity problems in software engineering. Moreover, the nodes of complex networks in these previous work were abstract concepts manually set by researchers and experts, leading to difficulties in building complex networks directly from data.

\vspace{0.5em}
\noindent Under such a background, we conducted further analysis of complexity problems on the collected dataset. Especially, we applied the complex network theory to software source code and modeled code tokens as network nodes for the first time, with the aspiration to call for the rethinking of complexity in software engineering.

\subsection{ Contributions of the Present Work}
Our work will contribute to related areas in mainly two aspects:

\begin{itemize}
    \item We proposed the first code \emph{dataset} mined from large-scale and industrial-strength source code. The dataset contains a massive amount of code data with complex relations, supporting an array of real-world software engineering tasks including code comprehension.
    \item By extracting code tokens and constructing relational networks, we pioneeringly incorporated complex network theories into software code representations, and created a novel and implementable \emph{methodology} to measure and address the complexity of source code in industrial settings.
\end{itemize}

\begin{figure*}
\centering
\includegraphics[width=0.8\linewidth]{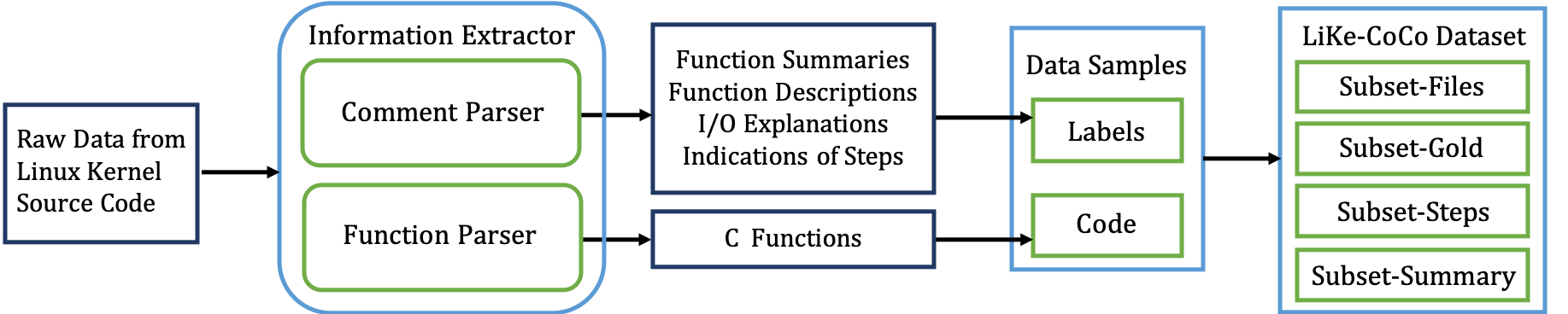}
\vspace{0.4em}
\nocaptionrule
\caption{\textbf{The dataset construction pipeline on Linux Kernel Code Repository}}
\label{fig2}
\end{figure*}

\section{Building Dataset on Linux Kernel Repository}
The dataset construction work on Linux Kernel source code involved two integral parts: automatic data collection and fit-for-purpose data partition (Figure 2 illustrates this framework).

\vspace{0.5em}
\noindent We extracted the comments and functions separately from the source code, and combined them to constitute samples in the dataset (see Figure 1 for example). For comment extraction, we applied the Comment Parser toolkit \cite{commentparser2015}, a Python module designed to extract comments from source code, to the code files of Linux kernel repository. For function extraction, we implemented a pattern matching algorithm to help recognize C functions' beginning and ending positions in code files. Syntax-related characters were reserved in the code data of each extracted functions for the further building of structural representations.

\vspace{0.5em}
\noindent Upon the information extraction on 12 critical folders in the repository, we constructed our LiKe-CoCo (\underline{Li}nux \underline{Ke}rnel \underline{Co}de \underline{Co}mprehension) Dataset, the scale of it amounts to nearly 140k C function samples. Table 1 shows the comprehensive statistics of LiKe-CoCo Dataset.

\vspace{0.5em}
\noindent According to the information contained in each data sample, data cleaning and partition were further performed. We established four subsets of the LiKe-CoCo Dataset to support various software engineering and code comprehension tasks:

\begin{itemize}
    \item The \textbf{LiKe-CoCo(files)} subset was
    designed for the mining of relations between functions in the range of a file, as well as the overall summarization of a code file. Each sample is a list of all the functions and corresponding comments within a C code file. 
    \item The \textbf{LiKe-CoCo(gold)} subset was established for multiple code comprehension tasks including function summarization, input/output tracking, and description generation. The samples are C functions with explicit and clearly-written comments before the function beginnings, indicating the summary, input/output description, and detailed description of the function. We view them as gold-quality labeled data. 
    \item The \textbf{LiKe-CoCo(steps)} subset was collected for the learning of internal processes within long function bodies, as well as the relations between these steps and the overall function. Each of the samples contains an extracted C function, a function summary, and some short comments inside the function body, serving as signs of internal steps towards function implementation. 
    \item The \textbf{LiKe-CoCo(sumry)} subset was built for the supervised learning of general function summarization tasks. There are only two integral components in a data sample: a C function and a function summary extracted from comments before the function beginning. 
\end{itemize}

\section{Model Source Code as Complex Network}
After dataset construction, we tokenized the C code within each sample. Relation parsing, network analysis, and knowledge management were implemented on the extracted tokens to address the complexity of code.

\subsection{Nodes of Network: Token Extraction}

\begin{table}
  \centering
  \nocaptionrule
  \caption{\textbf{The statistics of the overall dataset and four subsets} \\ (including the average length of code (in lines) in the dataset and the amount of samples from 12 folders of Linux kernel repository).}
    \vspace{0.5em}
    \begin{tabular}{lrrrrr}
    \toprule
          & \multicolumn{1}{l}{Overall} & \multicolumn{1}{l}{Files} & \multicolumn{1}{l}{Gold} & \multicolumn{1}{l}{Steps} & \multicolumn{1}{l}{Sumry} \\
    \midrule
    Avg Length & 23.04 & 23.04 & 27.67 & 44.32 & 28.64 \\
    Sample Num & 136428 & 8527  & 9617  & 16209 & 36686 \\
    \textit{from folders:} &       &       &       &       &  \\
    arch & 47940 & 4681  & 1505  & 4326  & 9683 \\
    block & 1936  & 70    & 349   & 248   & 615 \\
    crypto & 2004  & 150   & 73    & 168   & 352 \\
    certs & 16    & 4     & 2     & 2     & 10 \\
    fs & 29073 & 1277  & 2287  & 4851  & 10233 \\
    ipc & 306   & 11    & 33    & 42    & 78 \\
    kernel & 12168 & 353   & 1302  & 1693  & 4103 \\
    lib & 3269  & 341   & 520   & 284   & 927 \\
    mm & 4727  & 116   & 531   & 873   & 1775 \\
    net & 32184 & 1392  & 2196  & 3248  & 7532 \\
    security & 2494  & 125   & 808   & 452   & 1318 \\
    virt & 311   & 7     & 11    & 22    & 60 \\
    \bottomrule
    \end{tabular}%
  \label{tab:addlabel}%
\end{table}%

The code tokenization was conducted through Pygments, a generic syntax highlighting engine for source code \cite{Pygments2006}. Each C function in the dataset was transformed into a set of tokens, paired with token labels that explain corresponding tokens' roles. The extracted tokens could be categorized into four types: 

\begin{itemize}
    \item \textbf{Keywords} of C programming language, including keywords of control structures, such as "if", "for", "continue" and "break", as well as types of variables, such as "int" and "struct". The keyword tokens could help highlight the information flow in programs \cite{hendrix2002effectiveness}, playing a critical role in recognizing the \emph{structures} of code blocks.  
    \item \textbf{Names}, including function names, class names and variable names. The naming of variables and functions is an indispensable component in comprehending code \emph{semantics}, as the meaning of these names will directly influence our initial understanding of programs \cite{allamanis2015suggesting}.
    \item \textbf{Operators}, such as "=", "+", "!", and "\&". In addition to the name tokens, operator tokens also contribute to the \emph{semantics} of code blocks.
    \item \textbf{Punctuations}, including "(", ")", "\{", "\}", and ";". Punctuation tokens also help accentuate the \emph{structures} and relations within code tokens.
\end{itemize}

\begin{table*}
  \centering
  \nocaptionrule
  \caption{\textbf{The statistical analysis on tokenized code and network-based representations}}
  (including the average amount of various code tokens, as well as the mean value of maximum DC and MD in constructed networks)
  \vspace{0.5em}
    \begin{tabular}{lrrrrrrrr}
    \toprule
          & \multicolumn{1}{l}{Token Num} & \multicolumn{1}{l}{Keyw Num} & \multicolumn{1}{l}{Name Num} & \multicolumn{1}{l}{Punc Num} & \multicolumn{1}{l}{Operator Num} & \multicolumn{1}{l}{Node Num} & \multicolumn{1}{l}{Max DC} & \multicolumn{1}{l}{Mean Dist} \\
    \midrule
    Overall Dataset & 157.58 & 13.45 & 50.56 & 55.29 & 31.87 & 183.43 & 0.23  & 4.11 \\
    \textit{from folders:} &       &       &       &       &    &   &   & \\
    arch & 134.8 & 12.11 & 41.34 & 47.47 & 25.29 & 156.02 & 0.24  & 3.99 \\
    block & 136.44 & 11.58 & 44.3  & 47.2  & 29.07 & 158.06 & 0.23  & 4.15 \\
    crypto & 179.48 & 15.53 & 57.78 & 62.78 & 36.11 & 207.3 & 0.24  & 4.31 \\
    certs & 120.5 & 9     & 37.5  & 47    & 18    & 140   & 0.26  & 4.19 \\
    fs & 217.85 & 16.43 & 70.27 & 75.7  & 45.71 & 256.36 & 0.2   & 4.35 \\
    ipc & 139.91 & 11.21 & 46.39 & 47.09 & 32.7  & 159.91 & 0.21  & 4.02 \\
    kernel & 110.74 & 10.19 & 35.8  & 39.61 & 22.39 & 128.73 & 0.26  & 3.94 \\
    lib & 125.17 & 12.92 & 38.42 & 44.57 & 23.89 & 145.03 & 0.26  & 4.11 \\
    mm & 140.11 & 13.09 & 47.46 & 53.88 & 26.23 & 168.46 & 0.23  & 4.18 \\
    net & 155.79 & 13.28 & 51.05 & 54.29 & 31.91 & 180.02 & 0.22  & 4.07 \\
    security & 147.09 & 14.43 & 46.36 & 50.93 & 29.62 & 169.49 & 0.24  & 3.99 \\
    virt & 181.55 & 17.45 & 58.36 & 62.45 & 40.73 & 208.91 & 0.18  & 4.43 \\
    \bottomrule
    \end{tabular}%
  \label{tab:addlabel}%
\end{table*}%

\vspace{0.5em}
\noindent The detailed statistics of code tokens are included in Table 2. There are about 150 code tokens in each data sample, much more than the number of code tokens in the datasets\cite{wan2018improving, hu2018summarizing} for NLP-based code comprehension methods (nearly 50 tokens in each Python sample and 120 tokens in each Java sample). Punctuation, name, operator, and keyword tokens respectively occupy nearly 30\%, 30\%, 20\%, and 9\% of the total token amount. Taking these high ratios into account, the neglect of either structural or semantic information contained in the four types of tokens will lead to a consequential loss of comprehensive code understanding. 

\subsection{Edges within Network: Relation Parsing}

Taking the extracted tokens as nodes, we constructed network-based representations to represent the structure of each C function. The established structure networks are undirected and unweighted, with edges indicating the compositional relations between nodes. 

\vspace{0.5em}
\noindent To identify the network structure within a sequence of code tokens, we incorporated some interpretive nodes, such as "if statement" nodes and "while statement" nodes, to help segment the code tokens and describe the relations between nodes. When parsing code tokens, the relation parser assigned the corresponding interpretive node to a group of tokens within a code block based on its contained keywords and punctuations. The code token group was then recursively decomposed into lower-level groups of tokens, with more detailed interpretive nodes assigned to each group. Figure 3 has shown a clear example of this process.

\vspace{0.5em}
\noindent In contrast to the previous methods based on function call graph \cite{bohnet2006visual} and AST trees \cite{chilowicz2009syntax} to abstractly represent source code, this representation method will clearly reflect the internal structures of C functions, including control structures and long-term dependencies, without oversimplifying or discarding the rich semantics of code tokens. It provides us with a novel approach to leverage both structural and semantic information in code comprehension.

\subsection{Measure the Network: Complexity Evaluation}

\begin{figure}
\centering
\includegraphics[width=0.9\linewidth]{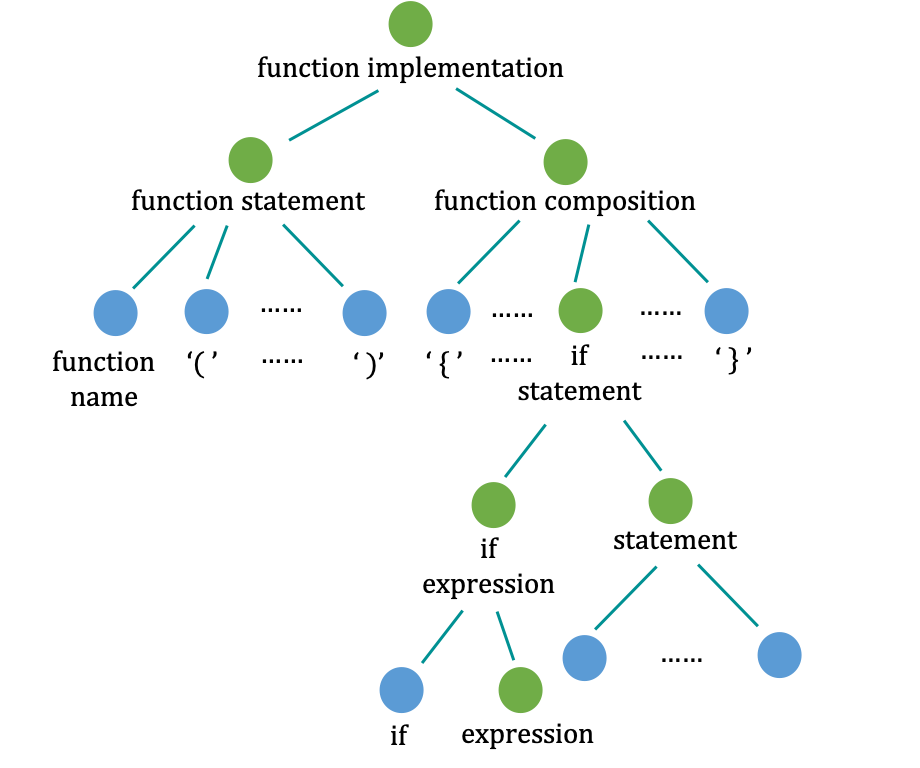}
\nocaptionrule
\caption{\textbf{An example of the network-based representations} \\ (The green nodes refer to interpretive nodes, while the blue nodes are representive of code tokens.)}
\label{fig3}
\end{figure}

\vspace{0.5em}
\noindent The large amount of nodes in network representations (see Table 2 for the statistics on network structures) has shown the complexity of source code in data samples to some degree. To systematically measure the structural complexity, we further computed several indicators \cite{lu2009analysis}:

\begin{itemize}
    \item To quantify the \textbf{density} of network representations, we focus on the \emph{degree} of nodes in structural graphs. For a complex network with $n$ nodes, the degree $k_i$ of a node $i$ refers to the number of edges related to it. A node's importance can be further evaluated through degree centrality (DC) by formula (1). As the degree centrality of nodes increases, the network becomes denser. Therefore, we measured the density of structure networks by the maximum DC in each sample. 
    \begin{equation}
        DC(i) = \frac{k_i}{n-1} \label{eq:DC}
    \end{equation}
    \item To evaluate the \textbf{scale} of structural networks, we considered the \emph{distances} between pairs of nodes. For a connected network with $n$ nodes, the distance $d_{ij}$ between node $i$ and node $j$ refers to the length of the shortest path between the two nodes. The mean distance (MD) computed by formula (2) indicates the sparsity between nodes. As the mean distance increases, the scale of the network becomes larger. To calculate this indicator, We implemented the Floyd algorithm\cite{wei2010optimized}, a classical algorithm in finding shortest paths, to compute the shortest distance between node pairs.
    \begin{equation}
        MD=\frac{\sum_{i \neq j}d_{ij}}{n(n-1)}\label{eq:MD}
    \end{equation}
\end{itemize}

\noindent Table 2 has shown the results of our computation. The degree centrality of nodes in each network exceeded the value of 0.2, which means there are some nodes associated with more than 1/5 of all the possible edges, reflecting the high density of relations within structure networks. The mean distances between nodes are also noteworthy. According to the algorithm output, there were averagely more than four edges in the shortest path between each pair of nodes in a network. Such a phenomenon indicates the sparsity and large scale of structure networks.

\subsection{Bridge Networks: Knowledge Base Settings}

While the network representations expressed the internal structure of C functions, a domain-specific knowledge base of Linux Kernel could further help us understand the token relations in the whole repository scope, thus bridging the networks established for individual data samples.

\vspace{0.5em}
\noindent The knowledge base mainly focused on name tokens. The occurrences of same name nodes in multiple networks were seen as the indications of network connection for complex networks and the signs of call relations in software source code. For each name token extracted from Linux kernel source code, we collected its meaning from related comments and the co-occurrence relations with other name tokens. The name, meaning, and relations were then reserved into a database as the domain-specific knowledge of Linux kernel. There currently are 50016 items in the knowledge base, and 9029 of them incorporated detailed descriptions from comments. The average number of co-occurrence relations associated with a name token is 85.31, reflecting the complex token connections within the whole Linux kernel repository.

\vspace{0.5em}
\noindent The established knowledge base will also largely facilitate the knowledge management of Linux kernel source code, thus improving the performance and interpretability of automatic code comprehension. Besides the efficient searching of token explanations, the knowledge base also assists us in the tracking of interleaved relations between tokens, leading to a deeper understanding of token semantics from related tokens, instead of the shallow interpretation from only the naming of tokens.

\section{Discussions}
Starting from code comprehension, we established the first code dataset collected from large-scale software repositories. Detailed analysis of code structures underlined the complexity of token relations in source code. The constructed network representations and knowledge base further help to evaluate and resolve the complexity in understanding code. Our findings and methodologies are promising for future research work on code comprehension, and highly meaningful to the overall process of software engineering.

\subsection{The Reach and Application of Our Work}
\subsubsection{Code Comprehension}
From the perspective of \textbf{Machine Code Comprehension}, the network-based representations of source code have brought great convenience and opportunities for future analysis and application based on the well-developed complex network theory. By analyzing the structure networks, downstream models will quickly recognize the quality, functionality, schema, and composition of source code. For example:
\begin{itemize}
    \item The detection of circuits in the constructed network \cite{chen2017robustness} could serve as a kind of evaluation for code robustness, as more circuits in the network means more redundant paths between code tokens.
    \item The identification of similar network structures \cite{zhang2018measure} will help the classification of coding schemas, thus facilitating the recognition of code functionalities.
    \item The analysis of network modularity \cite{newman2004fast, fortunato2007resolution} will assist in measuring how well the network will be partitioned into a group of sub-networks, providing a possible approach to tackle the complexity of structure networks by disintegration.
\end{itemize}

\vspace{0.5em}
\noindent
From the perspective of \textbf{Human Code Comprehension}, the data collection pipeline on source code will help people understand software code more efficiently. By building datasets and knowledge base on software code repositories, we will better discover and organize the knowledge contained in source code. For example, our dataset could help engineers and students systematically learn about operating systems and Linux kernel. Further applications of this pipeline to high-tech companies' leading products will also enable the establishment of internal datasets and knowledge bases for targeted code comprehension, largely alleviating engineers' working burden.

\subsubsection{More Than Code Comprehension}
\vspace{0.5em}
\noindent From the perspective of \textbf{Software Engineering Process}, our ideas and methods on complexity will facilitate almost all the components of software engineering, including the process of design, coding, debugging, testing, and sustaining, as well as the team management and communication in collaborative engineering \cite{Amy2016coop}. For example:
\begin{itemize}
    \item The analysis of code complexity on different versions of software source code will outline software evolution more clearly, bringing convenience for performance optimization, testing, and sustaining.
    \item The network-based representations could help the design of software API for different groups of people, such as customers and engineers, by emphasizing different network nodes.
    \item The complexity measurement and network representations on code will also facilitate the work assignment of a developers' team and the knowledge sharing between engineers, especially between experts and novices.
\end{itemize}

\subsection{Future Directions}
Our work will strengthen the research on incorporating structural information into the code comprehension process. Several previous works mainly focused on the performance improvement of neural network-based models by including code structure representations \cite{mou2016convolutional,ben2018neural,ahmad2020transformer}. We here highlighted the prospect of conducting code comprehension tasks directly on structural representations such as code structure networks, and exploiting neural network models' strength in processing big data to adjust parameters for the representations. Combining knowledge-based representations and data-driven learning will lead to more controllable, interpretable, and widely applicable comprehensions of source code.

\vspace{0.5em}
\noindent Our work will also inspire the modeling of the software engineering process based on complex network theory. We extended the application of complex networks to software source code for the first time and revealed its power in representing and quantifying code complexity for comprehension. Future research work needs to build more extensive networks on the whole source code repository, enable the network representations to learn from data, and simplify the network complexity for further computations and applications.

\section{Conclusion}
In the present work, we systematically collected a large-scale code dataset from the open-source Linux Kernel repository to support complex code comprehension and further software engineering tasks in industrial settings. Token extraction, network representations, and database construction strengthened the interpretation, abstraction, and application of code structures. The complex network-based analysis on code structure networks provided us with a novel approach to measure the complexity of code for comprehension tasks. Together, we call for further exploration of real-world code comprehension on software source code, comprehensive evaluation of industrial code complexity, and in-depth code understanding that combined both structural and semantic information.

\appendix


\acks

We are particularly thankful to Mr. Mingzhe Li for his contribution and assistance in the development of C code parsing tools in our work. We also thank Mr. Sitian Qian, Mr. Haoyu Wu, Mr. Haoren Deng, Mr. Tianyi Huang, and Mr. Wenlan Wei for their active participation in our seminars and discussions. Working with them has brought us many interesting ideas and research insights. We are also grateful to Dr. Frank Che, Dr. Ziyu Yao, Ms. Yajing Zhang, and Ms. Harper Liu. Their technical support and suggestions has helped us much in the scoping and solving of research problems. 


\bibliographystyle{abbrvnat}
\bibliography{references}

\end{document}